\documentclass[12pt]{article}
\usepackage{fancyhdr}

\usepackage{mathrsfs}
\usepackage[T1]{fontenc}
\usepackage{mathpazo}
\usepackage{setspace}
\usepackage{amsfonts}
\usepackage{amssymb}
\usepackage{amsmath}
\usepackage{epsfig}
\usepackage{latexsym}
\usepackage{color}
\usepackage{graphicx}
\usepackage{nicefrac}
\usepackage[latin1]{inputenc}
\usepackage{slashed}
\usepackage{multirow}
\usepackage{fancybox}
\usepackage{cite}
\usepackage{comment}
\usepackage{soul}
\usepackage{color}

\usepackage[all,cmtip]{xy}
\usepackage{hyperref}

\def\hybrid{\topmargin -20pt    \oddsidemargin 0pt
        \headheight 0pt \headsep 0pt
        \textwidth 6.25in       
        \textheight 9.25in       
        \marginparwidth .875in
        \parskip 5pt plus 1pt   \jot = 1.5ex}

\hybrid

\def\baselinestretch{1.2}

\catcode`\@=11

\def\marginnote#1{}
\newcount\hour
\newcount\minute
\newtoks\amorpm
\hour=\time\divide\hour by60
\minute=\time{\multiply\hour by60 \global\advance\minute by-\hour}
\edef\standardtime{{\ifnum\hour<12 \global\amorpm={am}%
        \else\global\amorpm={pm}\advance\hour by-12 \fi
        \ifnum\hour=0 \hour=12 \fi
        \number\hour:\ifnum\minute<10 0\fi\number\minute\the\amorpm}}
\edef\militarytime{\number\hour:\ifnum\minute<10 0\fi\number\minute}

\def\draftlabel#1{{\@bsphack\if@filesw {\let\thepage\relax
   \xdef\@gtempa{\write\@auxout{\string
      \newlabel{#1}{{\@currentlabel}{\thepage}}}}}\@gtempa
   \if@nobreak \ifvmode\nobreak\fi\fi\fi\@esphack}
        \gdef\@eqnlabel{#1}}
\def\@eqnlabel{}
\def\@vacuum{}
\def\draftmarginnote#1{\marginpar{\raggedright\scriptsize\tt#1}}

\def\draft{\oddsidemargin -.5truein
        \def\@oddfoot{\sl preliminary draft \hfil
        \rm\thepage\hfil\sl\today\quad\militarytime}
        \let\@evenfoot\@oddfoot \overfullrule 3pt
        \let\label=\draftlabel
        \let\marginnote=\draftmarginnote
   \def\@eqnnum{(\theequation)\rlap{\kern\marginparsep\tt\@eqnlabel}%
\global\let\@eqnlabel\@vacuum}  }

\def\preprint{\twocolumn\sloppy\flushbottom\parindent 2em
        \leftmargini 2em\leftmarginv .5em\leftmarginvi .5em
        \oddsidemargin -.5in    \evensidemargin -.5in
        \columnsep .4in \footheight 0pt
        \textwidth 10.in        \topmargin  -.4in
        \headheight 12pt \topskip .4in
        \textheight 6.9in \footskip 0pt
        \def\@oddhead{\thepage\hfil\addtocounter{page}{1}\thepage}
        \let\@evenhead\@oddhead \def\@oddfoot{} \def\@evenfoot{} }

\def\numberbysection{\@addtoreset{equation}{section}
        \def\theequation{\thesection.\arabic{equation}}}

\def\underline#1{\relax\ifmmode\@@underline#1\else
        $\@@underline{\hbox{#1}}$\relax\fi}

\def\titlepage{\@restonecolfalse\if@twocolumn\@restonecoltrue\onecolumn
     \else \newpage \fi \thispagestyle{empty}\c@page\z@
        \def\thefootnote{\fnsymbol{footnote}} }

\def\endtitlepage{\if@restonecol\twocolumn \else \newpage \fi
        \def\thefootnote{\arabic{footnote}}
        \setcounter{footnote}{0}}  

\catcode`@=12
\relax

\def\figcap{\section*{Figure Captions\markboth
        {FIGURECAPTIONS}{FIGURECAPTIONS}}\list
        {Figure \arabic{enumi}:\hfill}{\settowidth\labelwidth{Figure
999:}
        \leftmargin\labelwidth
        \advance\leftmargin\labelsep\usecounter{enumi}}}
 \relax
\def\tablecap{\section*{Table Captions\markboth
        {TABLECAPTIONS}{TABLECAPTIONS}}\list
        {Table \arabic{enumi}:\hfill}{\settowidth\labelwidth{Table
999:}
        \leftmargin\labelwidth
        \advance\leftmargin\labelsep\usecounter{enumi}}}
 \relax
\def\reflist{\section*{References\markboth
        {REFLIST}{REFLIST}}\list
        {[\arabic{enumi}]\hfill}{\settowidth\labelwidth{[999]}
        \leftmargin\labelwidth
        \advance\leftmargin\labelsep\usecounter{enumi}}}
 \relax
\makeatletter
\newcounter{pubctr}
\def\publist{\@ifnextchar[{\@publist}{\@@publist}}
\def\@publist[#1]{\list
        {[\arabic{pubctr}]\hfill}{\settowidth\labelwidth{[999]}
        \leftmargin\labelwidth
        \advance\leftmargin\labelsep
        \@nmbrlisttrue\def\@listctr{pubctr}
        \setcounter{pubctr}{#1}\addtocounter{pubctr}{-1}}}
\def\@@publist{\list
        {[\arabic{pubctr}]\hfill}{\settowidth\labelwidth{[999]}
        \leftmargin\labelwidth
        \advance\leftmargin\labelsep
        \@nmbrlisttrue\def\@listctr{pubctr}}}
 \relax
\makeatother
\newskip\humongous \humongous=0pt plus 1000pt minus 1000pt

\newif\ifdtup

\relax

\def\be{\begin{equation}}
\def\ee{\end{equation}}
\def\ba{\begin{eqnarray}}
\def\ea{\end{eqnarray}}

\def\no{\noindent}

\def\IR{\relax{\rm I\kern-.18em R}}

\def\IR{\relax{\rm I\kern-.18em R}}
\def\IL{\relax{\rm I\kern-.18em L}}

\def\inv{^{\raise.15ex\hbox{${\scriptscriptstyle -}$}\kern-.05em 1}}

\def\beq{\begin{equation}}
\def\eeq{\end{equation}}
\def\bea{\begin{eqnarray}}
\def\eea{\end{eqnarrat}}

\begin{document}

\renewcommand{\theequation}{\thesection.\arabic{equation}}
\csname @addtoreset\endcsname{equation}{section}

 \def\baselinestretch{1.2}
 \noindent

\begin{titlepage}
\begin{center}


~

\vskip 1 cm

{\large \bf  Generalised T-duality and Integrable Deformations }

\vskip 0.4in

 {\bf Daniel C. Thompson}$^1$\vskip 0.1in
{\em
{\it ${}^1$ Theoretische Natuurkunde, Vrije Universiteit Brussel,\\
\& The International Solvay Institutes,\\
Pleinlaan 2, B-1050 Brussels, Belgium.\\ 
\vskip
0.1in
 }{\tt \footnotesize daniel.thompson@vub.ac.be}}
\vskip .5in
\end{center}

\centerline{\bf Abstract}
\no
We review some recent developments in the construction of integrable  $\eta$- and $\lambda$-deformations of the $AdS_5 \times S^5$ superstring. We highlight their link with Poisson-Lie T-duality.    Proceedings of a talk at ``The String Theory Universe, $21^{st}$ European String Workshop and $3^{rd}$ COST MP1210 Meeting''.

\no

\newpage


\noindent

\vskip .4in

\end{titlepage}
\vfill
\eject

\def\baselinestretch{1.2}
\baselineskip 20 pt
\noindent
\def\jr{}
\section{Introduction}

Integrability has been a game changer in the context of AdS/CFT over the past decade and gives previously unfeasible calculational power.  The formidable technical machinery of integrability can sometimes obscure the true message -- integrability is really a statement that a physical system is simple.  In holography the classic example is that  the dimensions of operators  of ${\cal N}=4$ Super Yang-Mills (SYM) can be mapped to energy levels of an integrable spin chain  \cite{Minahan:2002ve}. That integrability emerges in this way in a 4d gauge theory is remarkable. 

One might wonder whether this integrability was just a consequence of the maximally (super)symmetric setting of ${\cal N}=4$  SYM.   A very natural question we are led to ask is if we can relax some of the assumptions of (super)symmetry whilst preserving integrability?  One reason for such hope   is that  even QCD itself exhibits some integrability  in certain limits \cite{Faddeev:1994zg,Lipatov:1994xy}. ${\cal N}=4$ SYM   admits a class of marginal deformations  known as (real) $\beta$-deformations, that modify the superpotential, preserve only the minimal amount of (conformal) supersymmetry but yet preserve integrability.  From the holographic perspective  the geometry describing these $\beta$-deformations  can be obtained by the application of $T$-dualities in TsT transformations \cite{Lunin:2005jy}.    Remarkably the string   worldsheet in the $\beta$-deformed  spacetime remains integrable  \cite{Frolov:2005dj}. 

The past two years have seen the development of two new classes of integrable deformations to the $AdS_5\times S^5$ superstring called $\eta$- \cite{Delduc:2013fga,Delduc:2013qra} and $\lambda$-deformations \cite{Sfetsos:2013wia,Hollowood:2014qma}.  At first sight these deformations are even more severe than the $\beta$-deformation; they preserve {\em no} supersymmetries, $\eta$-deformations preserve only some $U(1)$ isometries and $\lambda$-deformed spacetimes have no isometries at all.  However, it has been conjectured that the deformed theories actually realise the missing symmetries in the sense of a quantum group.  The S-matrix of the $\eta$ and $\lambda$ deformed $AdS_5\times S^5$ superstring is expected to be the unique quantum group deformed S-matrix  \cite{Hoare:2011wr} with the quantum group parameter real in the case of $\eta$-deformations \cite{Delduc:2013qra} and a root of unity in the case of $\lambda$ \cite{Hollowood:2014qma}.  

These deformations provide a new angle on the rich interplay of holography, integrability and T-duality.  We will see that though we derive the $\eta$ and $\lambda$ deformed theories two very different ways, they are related by a combination of analytic continuation  and generalised T-dualities.  The type of T-duality involved is known as a Poisson-Lie T-duality \cite{Klimcik:1995ux} and provides a notion of T-duality in a target space without isometries.    

In this short talk (and proceedings) working with the entire $AdS_5 \times S^5$ superstring in a super-coset formulation would be rather unedifying so for pedagogical purpose we illustrate the developments in a simpler scenario. For the $\eta$-deformation we will consider a deformation of the Principal Chiral Model (PCM) of bosonic strings on a group space first introduced by Klimcik  \cite{Klimcik:2002zj}  by the name of Yang-Baxter deformations.  For the $\lambda$-deformation we will deform a bosonic Wess-Zumino-Witten (WZW) model as in \cite{Sfetsos:2013wia}.

\section{$\eta$-Deformed Principal Chiral Model}

Let us begin with a toy model; the PCM on the three-sphere.  This is a theory of maps $g:\Sigma \rightarrow SU(2)$ with an action, 
\begin{equation}
\label{eq:PCM}
S_{PCM}[g]= \frac{\kappa^2}{4\pi} \int_\Sigma {\textrm{d}}^2 \sigma {\textrm{Tr}}  \left( g^{-1} \partial_+ g g^{-1} \partial_- g \right)\, .
\end{equation}
Bosonic PCMs such as this are not conformal but can appear as a subsector of a full string theory. Moreover, they are interesting in their own right since, as asymptotically free two-dimensional QFTs, they serve as good toy models to understand features of QCD. 

The case at hand, Eq.~\eqref{eq:PCM}, possesses a global $SU(2)_L\times SU(2)_R$  symmetry and its dynamics are classically integrable; the equations of motions and Bianchi identities for the currents, $J^a  =  {\textrm{Tr}} T^a  g^{-1}{\textrm{d}} g$, can be written as a flatness condition for a   Lax connection, 
\begin{equation}
\label{eq:Lax}
L(z) = \frac{1}{1-z^2} J + \frac{z}{1-z^2} \star J \, , \quad {\textrm{d}}L - L\wedge L = 0  \, . 
\end{equation}
From this Lax connection an $\infty$ tower of non-local conserved charges can be generated from   $T(z) = P \exp \int d\sigma L_\sigma$.

It has been known since the work of Cherednik \cite{Cherednik:1981df} that this theory admits an integrable deformation,
\begin{equation}
\label{eq:PCMdef}
S_C[g]= \frac{\kappa^2}{4\pi} \int_\Sigma {\textrm{d}}^2 \sigma {\textrm{Tr}}  \left( g^{-1} \partial_+ g g^{-1} \partial_- g \right) +  C J_+^3 J_-^3 \, .
\end{equation}
This action defines a non-linear $\sigma$-model into a squashed three-sphere target space and whilst still integrable, the global symmetries are broken down to $SU(2)_L \times U(1)_R$.   What is rather amazing \cite{Kawaguchi:2011pf} is that the broken symmetry is recovered from the non-local charges in the form of a semi-classical realisation of the quantum group ${\cal U}_q(\frak{sl}_2)$ which begins with, 
\begin{equation}
\label{eq:QG}
\{ Q^+_R , Q_R^- \}_{P.B.} = - i \frac{q^{Q^3_R} - q^{-Q^3_R} }{q - q^{-1}} \, , \quad q = \textrm{exp}\left(\frac{\sqrt{C}}{1+ C} \right) \, .
\end{equation}

This idea can be generalised to the PCM on arbitrary groups, $G$, following the work of Klimcik  \cite{Klimcik:2002zj}. A central r\^ole is played by a ${\cal R}$-matrix that solves a {\em modified} Yang-Baxter equation,
\begin{equation}
\label{eq:mYB}
[{\cal R} A, {\cal R} B] - {\cal R}([{\cal R} A,B]+[A,{\cal R} B] ) = [A, B]  \, , 
\end{equation}
which should hold for all $A,B \in \frak{g}$ and where the modification is the presence of the term on the right hand side. 
For compact bosonic groups such a solution is unique e.g. for the case of $\frak{su}(2)$ one has ${\cal R}: ( \sigma^1, \sigma^2, \sigma^3) \mapsto (- \sigma^2 , \sigma^1, 0)$.  Armed with this we defined the $\eta$-deformed PCM  \cite{Klimcik:2002zj},
\begin{equation}
\label{eq:Seta}
 S_\eta[g] = \frac{1}{2\pi t} \int_\Sigma {\textrm{d}}^2 \sigma {\textrm{Tr}}  \left( g^{-1} \partial_+ g, \frac{1}{1-\eta {\cal R}} g^{-1} \partial_- g \right)  \, .
\end{equation}
This theory is integrable and its Lax connection can be explicitly written down.    As with the $SU(2)$ example above the global $G_R$ action is broken but is recovered through the Poisson algebra of non-local charges as a quantum group ${\cal U}_q(\frak{g})$ with $q =\textrm{exp}(\eta t \pi)$    \cite{Delduc:2013fga}.  

The work of Delduc {\em et al} extends this construction to first accommodate symmetric cosets  \cite{Delduc:2013fga} and eventually semi-symmetric spaces (i.e. super-cosets)   \cite{Delduc:2013qra} and in particular to the $PSU(2,2|4)$ case relevant for the $AdS_5\times S^5$ superstring.  To gain some flavour for the $(AdS_5\times S^5)_\eta$ deformed spacetime we note that the isometries are broken down to a $U(1)^6$ and that supersymmetry is not preserved. 


Based on the semi-classical quantum group symmetries of this theory it is conjectured \cite{Delduc:2013qra} that this provides a Lagrangian description for the quantum group deformation of the  $AdS_5\times S^5$ S-matrix   \cite{Hoare:2011wr} in  the case where the quantum group parameter is real.  A substantial piece of evidence in support of this has been the perturbative matching in the large tension limit \cite{Arutyunov:2013ega}.  

There remain a few important unresolved issues the most pressing  of which is to establish more firmly the status of the $\eta$-deformed $PSU(2,2|4)$ superstring as a true CFT.  As a first step one would like to show that the metric of the $(AdS_5\times S^5)_\eta$ deformed spacetime can be embedded into a full type IIB supergravity background. However  attempts in this direction have thus far been somewhat inconclusive and indeed rather puzzling \cite{Lunin:2014tsa,Arutyunov:2015qva}.

  {\em{Note Added}}: Recently further considerations of the supergravity interpretation have been provided in \cite{Arutyunov:2015mqj} in which it is suggested that "background" of the $\eta$-model is {\em not} a type IIB supergravity solution however the 2d worldsheet is UV finite and scale invariant.  Surprisingly, the background does solve some rather natural modified supergravity equations.  The meaning of this, and whether the $\eta$-model really defines a critical string theory  is a focus of current efforts.

\section{$\lambda$-Deformed WZW Model}

Now we turn to what will at first seem like a completely different construction making use of an idea introduced by Sfetsos  \cite{Sfetsos:2013wia}.  This simple recipe to construct integrable models goes as follows:  
\begin{enumerate}
\item {\bf Double}--  First we shall double the degrees of freedom by combining two integrable theories namely the PCM with radius $\kappa$ as defined in Eq.~\eqref{eq:PCM} and the WZW model at level $k$.  
\item{\bf Gauge}-- To reduce back to the original number of degrees of freedom we gauge the $G_L$ action in the PCM and the anomaly free diagonal action of $G$ in the WZW with a common gauge field.  Though the gauge fields are non-propagating in two-dimensions this ties the two integrable systems together.
\item{\bf Fix}-- We now make a gauge fixing choice  such that all that is left of the gauged PCM is a quadratic term in the gauge fields. 
\item{\bf Integrate Out}-- Now the gauge fields can be integrated out to result in a deformation of the WZW model. 
\end{enumerate}
This procedure, which somewhat resembles a Buscher T-dualisation, defines the $\lambda$-deformed theory with action, 
\begin{equation}
\label{eq:Slambda}
S_\lambda[g]= S_{WZW}[g] + \frac{k}{2\pi}\int {\textrm{Tr}}  \left( g^{-1} \partial_+ g, \frac{1}{\lambda^{-1} - \textrm{Ad}_g }  \partial_- g g^{-1} \right) \, , 
\end{equation}
where the parameter $\lambda$ measures the ratio of the PCM radius to WZW level,
\begin{equation}
\lambda = \frac{k}{\kappa^{2}+ k } \ . 
\end{equation}
The result remains integrable!  

To understand this $\lambda$-deformed model let us  first consider the limit of $\lambda\rightarrow 0$   which yields a deformation of a WZW model by a marginally relevant current bilinear,
\begin{equation}
S_{\lambda} |_{\lambda \rightarrow 0} \approx k S_{WZW} + \frac{k}{\pi}\int \lambda J_{+}^{a} J_{-}^{a} + {\cal O}( \lambda^{2} ) \ . 
\end{equation}
On the other hand upon taking the limit $\lambda \rightarrow 1$ (which must be done with some care) one finds that the gauged WZW model degenerates into a Lagrange multiplier enforcing a flat connection. Then integration out of gauge fields is now exactly the T-duality Buscher procedure applied to a  non-Abelian group of isometries. So in this limit--as was shown in \cite{Sfetsos:2013wia}--we recover the non-Abelian T-dual of the PCM,
 \begin{equation}
S_{\lambda} |_{\lambda \rightarrow 1} \approx \frac{1}{\pi} \int \partial_{+} X^{a} (\delta_{ab} + f_{ab}{}^{c}X_{c})^{-1} \partial_{-} X^{b}  + {\cal O}(k^{-1}) \ . 
\end{equation}
This result was my initial interest in the subject;  one can think that the $\lambda$-deformation gives some sort of `regularised' version of non-Abelian T-duality.  

It was rather quickly realised how to generalise this $\lambda$-deformations to symmetric cosets \cite{Hollowood:2014rla} and semi-symmetric super-cosets (for which this gauging procedure gives the correct bosonic sector but needs some amendment in the fermionic directions)\cite{Hollowood:2014qma}.   For the case based on the $PSU(2,2|4)$  super-coset, it was conjectured \cite{Hollowood:2014qma} that like $\eta$, this  $\lambda$-deformation   also provides   a Lagrangian realisation of a quantum group deformed S-matrix but with the parameter $q$ this time a root unity  $q = e^{i \frac{\pi}{k} }$.   Indeed, this is supported by the underlying structure of non-local charges in these models  \cite{Itsios:2014vfa},\cite{Hollowood:2015dpa}. 

As with the case of $\eta$ deformations we should like to understand if these theories are conformal (at least at one-loop).  The first observation is that the for a compact group the $\lambda$ deformation is a marginally relevant operator however for a non-compact group one has a marginally irrelevant operator.  Thus if we combine  a $\lambda$ deformation for a compact group with that of a non-compact group one has some hope that the two may provide equally and opposite cancelling contributions to the dilaton $\beta$-function.  This was show in \cite{Sfetsos:2014cea} for some low dimensional examples.  Further it was shown that these $\lambda$-deformations can be given embeddings as complete ten-dimensional type IIB supergravity solutions for cases relevant to $AdS_3\times S^3 \times  T^4$  \cite{Sfetsos:2014cea} and $AdS_5 \times S^5$ \cite{Demulder:2015lva}.  In this approach a simple form for the RR sector fields is ``boot-strapped'' from just knowledge of the bosonic NS sector  using techniques developed for non-Abelian T-duality in \cite{Sfetsos:2010uq}.  A more direct approach is to compute from first principles the one-loop $\beta$ function for the coupling $\lambda$.  This was done for the bosonic case in \cite{Itsios:2014lca} and for the supercoset case in \cite{Appadu:2015nfa}.  In fact the answer is very simple,
\begin{equation}
\mu \frac{\mathrm{d}\lambda  }{ \mathrm{d} \mu  } =  \frac{c_2( G)}{k} \lambda \,  ,
\end{equation} 
where we note $c_2(G)$, the quadratic Casimir in the adjoint, vanishes for $PSU(2,2|4)$.

A further recent line of development in this direction has been the construction of multi-parameter integrable $\lambda$-deformations \cite{Sfetsos:2015nya} in which $\lambda$ entering into the action Eq.~\eqref{eq:Slambda} is promoted to a matrix $\lambda_{ab}$.

\section{The $\eta$-$\lambda$ connection}

Now lets put $\eta$ and $\lambda$ together.  Though their Lagrangian construction is quite different  they are actually closely related via a generalisation of T-duality called Poisson-Lie (PL) T-duality  introduced in  \cite{Klimcik:1995ux} and used in the present context in \cite{Vicedo:2015pna,Hoare:2015gda,Sfetsos:2015nya,Klimcik:2015gba}.
 
To explain this  remember that the $\eta$-deformations broke the $G_{R}$ symmetry.  So the corresponding  currents ${\cal J}_{a}$--which are dressed versions of the Maurer-Cartan forms--are no-longer conserved. However the algebraic structure is very delicate and means that these currents obey a modified conservation law of the form, 
\begin{equation}
d\star {\cal J}_{a} = \tilde{f}^{bc}{}_{a}   {\cal J}_{b} \wedge   {\cal J}_{c} \, .
\end{equation}

If the right-hand side was zero, we could simply perform T-duality by gauging these currents.   It turns out even with this non-zero right hand side there is still a notion of a T-dual   introduced in the late 1990's by Klimcik and Severa \cite{Klimcik:1995ux}.  What makes it work is that  the $\tilde{f}$ are very special; they are structure constants for a second Lie-algebra $\tilde{\frak{g}}$ built using the Yang-Baxter ${\cal R}$-matrix to define a bracket,
\begin{equation}
[ A , B]_{{\cal R}} = [{\cal R} A , B]- [A,{\cal R} B]\, .
\end{equation}
Mathematically speaking, the sum  $ \frak{g} \oplus \tilde{\frak{g}} \cong \frak{g}^{\mathbb{C}}$ defines a Drinfel'd Double equipped with an inner product given in an obvious notation by,    
\begin{equation}
  \langle T_a , T_b \rangle = \langle \tilde T^a , \tilde T^b \rangle= 0 \ , \quad   \langle T_a , \tilde T^b \rangle = \delta_a^b\, .
 \end{equation}

PL T-duality    is an equivalence between   two $\sigma$-models,
 \begin{equation}
\begin{aligned}
&S[g] =\frac{1}{2\pi t\,\eta} \int  \mathrm{d}^2 \sigma J_+^T (M - \Pi )^{-1} J_-\,, \quad g \in \frak{g} \, ,  \nonumber \\
&\tilde{S}[\tilde{g}] =\frac{1}{2\pi t\,\eta}  \int \mathrm{d}^2 \sigma \tilde{J}_+^T (M^{-1} - \tilde\Pi )^{-1} \tilde{J}_-\, , \quad  \tilde g \in \frak{\tilde g} \, .\nonumber 
  \end{aligned}
\end{equation}
where $M$ can be an arbitrary constant matrix and the group theoretic matrix $\Pi$ is defined via, 
\begin{equation} 
 a_a{}^b = \langle g^{-1} T_a g , \tilde{T}^b \rangle \ , \quad  b^{ab} = \langle g^{-1} \tilde T^a g , \tilde{T}^b \rangle \ , \quad \Pi = b^T a\, ,
\end{equation}
 with similar for $\tilde\Pi$.   
 
 The $\lambda$-deformation takes exactly the form of the first of these PL dual pairs with the identification  $ M= \eta^{-1} - {\cal R} $ and hence can be T-dualised in this fashion.  This is not quite the $\lambda$-theory; it needs to be supplemented by analytic continuation of some of the Euler angles defining the group element and also a relation between the deformation parameters\footnote{In comparison to other instances in the literature one needs to take some care, here we have inherited definitions of $\lambda$ and $\eta$ that are most natural from bosonic $\sigma$-model point of view but in the context of superstrings  different definitions $\lambda_s$ and $\eta_s$ are more natural and sometimes used.  These are related by $\lambda = \lambda_s^2$  and $\eta = \frac{2 \eta_s}{1-\eta_s^2}$.     }  and tensions of the models,
\begin{equation}
 \eta \rightarrow \frac{i (1- \lambda)}{(1+\lambda)}, \quad t \rightarrow \frac{(1+\lambda)}{ k(1-\lambda)} \ . 
\end{equation}
Doing the combination of PL duality, analytic continutations and applying this relation maps exactly the $\eta$ theory of Eq.~\eqref{eq:Seta} to the $\lambda$ theory of Eq.~\eqref{eq:Slambda}.  Notice that acting on the parameter $q$ we have, 
\begin{equation}
q = e^{\eta t \pi} \leftrightarrow q = e^{\frac{i\pi}{k}} \, , 
\end{equation}
showing indeed $q$ real gets mapped to $q$ root unity. 
 
\section{Outlook}

To summarise, we've seen two exciting new classes of integrable models, the $\eta$ and the $\lambda$ deformations. They are connected by generalised T-dualities.   Their symmetries include quantum groups for the case of $q$ a real and $q$ a root unity respectively.   These approaches extend to superstrings in $AdS_{5}\times S^{5}$ and are opening a new window onto integrable deformations in the AdS/CFT conjecture.\\

There are many nice angles still to study:
\begin{itemize}  
\item What are the implications of Poisson-Lie duality for  Double Field Theory?
\item Are these true consistent CFTs?
\item Can this be extended to new scenarios e.g. $AdS_{4}\times CP^{3}$? 
\item Most important; and most difficult; what are the implications and interpretation of this the on gauge theory side?\\
\end{itemize}

{\em{I would like to thank the organisers and supporters of the conference and my collaborators on this topic  K. Sfetsos and K. Siampos.   Due to limitations of space in these proceedings this is a far from exhaustive bibliography and we apologise to the authors of the many excellent papers that are omitted. }}

\end{document}